\documentclass[11pt,authoryear,review,3p,times]{elsarticle}
\usepackage[english]{babel}
\usepackage[utf8]{inputenc}
\usepackage[T1]{fontenc}
\usepackage{graphicx,array,bm,color,natbib,amssymb}
\newtheorem{theorem}{Theorem}
\newdefinition{definition}{Definition}
\newproof{proof}{Proof}
\graphicspath{{figures/}{../figures/}}
\usepackage{subcaption}
\usepackage{float}
\journal{to the chosen journal}

\begin{document}
	
	\begin{frontmatter}
		\title{{ \LARGE Estimation of Component Reliability in Coherent Systems}}

		\author[label2]{\large Agatha S. Rodrigues\corref{aaa}}	
		\author[label2]{\large 	Felipe~Bhering}	
		\author[label2]{\large Carlos~Alberto~de~Bragan\c{c}a~Pereira}	
		\author[label6]{\large Adriano Polpo}
		
		\address[label2]{Institute of Mathematics and Statistics, University of S\~ao Paulo,  S\~ao Paulo, SP, Brazil.}

		\address[label6]{Department of Statistics, Federal University of S\~ao Carlos, S\~ao Paulo, SP, Brazil.}

		\cortext[aaa]{e-mail: agatha@ime.usp.br\\ \indent\hspace{.15cm} phone number: +55 11 986667332}
	

	\begin{abstract}
The first step in statistical reliability studies of coherent systems is the estimation of the reliability of each system component. For the cases of parallel and series systems the literature is abundant. It seems that the present paper is the first that presents the general case of component inferences in coherent systems.  The failure time model considered here is the three-parameter Weibull distribution. Furthermore, neither independence nor identically distributed failure times are required restrictions. The proposed model is general in the sense that it can be used for any coherent system, from the simplest to the more complex structures. It can be considered for all kinds of censored data; including interval-censored data. An important property obtained for the Weibull model is the fact that the posterior distributions are proper, even for non-informative priors. Using several simulations, the excellent performance of the model is illustrated. As a real example, boys’ first use of marijuana is considered to show the efficiency of the solution even when censored data occurs.
	\end{abstract}

 \end{frontmatter}

	%

	\newpage
	\section{Introduction}
	\label{secao1}
	
As motivation, the reliability estimation of an automatic coffee machine is described.  The machine has two possible causes of failure: 1. the failure of the component that grinds the grain; or 2. the failure of the water heating component.  Clearly, the failure of either, 1 or 2, leads the coffee machine failure. Here, the failure of a component implies that the possible future failure time of the other becomes invisible, i.e. a censored data. Statistical inference for the reliability of the machine depends on both marginal components probability models, even in the presence of censoring. Hence, inferences for both components are needed.

Statistical inference of component reliability is not an easy task: censoring, dependence and unequal distributions are some of the troubles.  Considering a sample of the coffee machine example for which all $n$ sample units are observed up to death, every sample unit will produce a component failure time and a censored failure time for the other component.  Both components failing at the same time is considered unlikely in such situations. In this example, the sample will produce $n$ failure time observations and $n$ censored times for the two components in the test. Relative to component failure time, it is reasonable to say that the two types of components are not identically distributed; it is likely that one of the component types may suffer more censoring than the other. It is common that only one component is responsible for the system failure at time $t$, implying that all the remaining components are censored also at time $t$, although the types of censoring could be different. The number of censored data may be much higher than the number of uncensored ones!

\begin{figure}[H]\centering
	\begin{minipage}[H]{0.25\linewidth}
		\includegraphics[width=3.5cm,height=1cm]{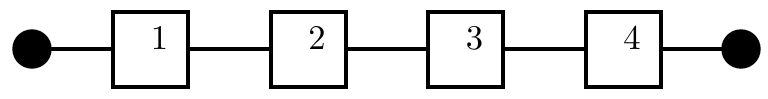}
		\subcaption{ $\mbox{   }$ }\label{series}
	\end{minipage} 
	\begin{minipage}[H]{0.15\linewidth}
		\includegraphics[width=2cm,height=2cm]{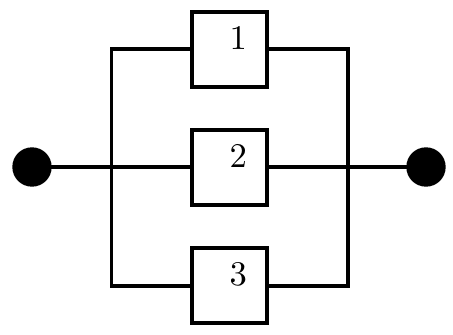}
		\subcaption{ $\mbox{   }$ }\label{parallel}
	\end{minipage}
	\caption{(a) Representation of a series system with $4$ components; (b) Representation of a parallel system with $3$ components.}
\end{figure}

The reliabilities of a system and its components also depend on the structure of the system, the way components are interconnected. The coffee machine is a series system of two components, a simple case known as competing risks problem. Figure \ref{series} is a series system of four components - at the time the system fails only one component is uncensored and the other three components are right-censored at the system failure time. A parallel system as in Figure \ref{parallel} works whenever at least one component is working. Again, only one component has its failure time uncensored; the other components are left-censored observations.

The literature on reliability of either parallel or series systems is abundant; different solutions have been presented. \citet{VHS1997}, \citet{SalinasTorres}, \citet{PolpoCarlinhos} and \citet{PolpoSinha} discussed the Bayesian nonparametric statistics for series and parallel systems. Under Weibull probability distributions, Bayesian inferences for system and component reliabilities were introduced by \citet{CoqueJr} and \citet{Bhering} presented a hierarchical Bayesian Weibull model for component reliability estimation in series and parallel systems, proposing a useful computational approach. Using simulation for series systems, \citet{AgathaIC}, considering Weibull families, compared three estimation types: Kaplan-Meier, Maximum Likelihood and Bayesian Plug-in Estimators. \citet{PolpoSinhaSimoniCAB} performed a comparative study about Bayesian estimation of a survival curve. 

\begin{figure}[H]\centering
	\begin{minipage}[H]{0.26\linewidth}
		\includegraphics[width=\linewidth]{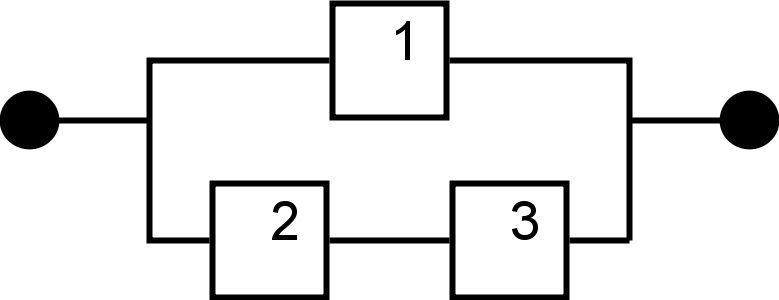} 
		\subcaption{ $\mbox{   }$ } \label{PSS}
	\end{minipage}
	\begin{minipage}[h]{0.28\linewidth}
	\includegraphics[width=\linewidth]{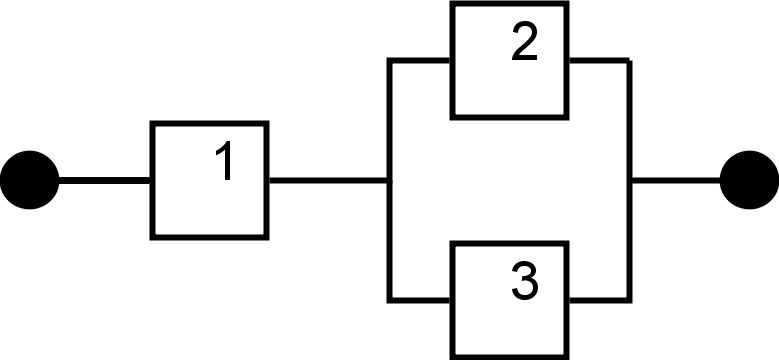}
	\subcaption{ $\mbox{   }$ }
\end{minipage} 
	\caption{(a) Parallel-series representation. (b) Series-parallel representation .}
	\label{systems_sps_pss}
\end{figure}

Considering the celebrated property that any coherent systems can be written as a combination of parallel and series systems, \citet{PolpoSiCar} introduced Bayesian nonparametric statistics for a class of coherent system. Figure \ref{systems_sps_pss} illustrates two cases of this kind of combinations with three components; component 2, for example, is susceptible to both right- and left-censoring. 
 \citet{PolpoSiCar} restricted themselves to cases for which no component appears more than twice in parallel-series and series-parallel representations.

\begin{figure}[H]\centering
	\centering
	\includegraphics[width=0.4\linewidth]{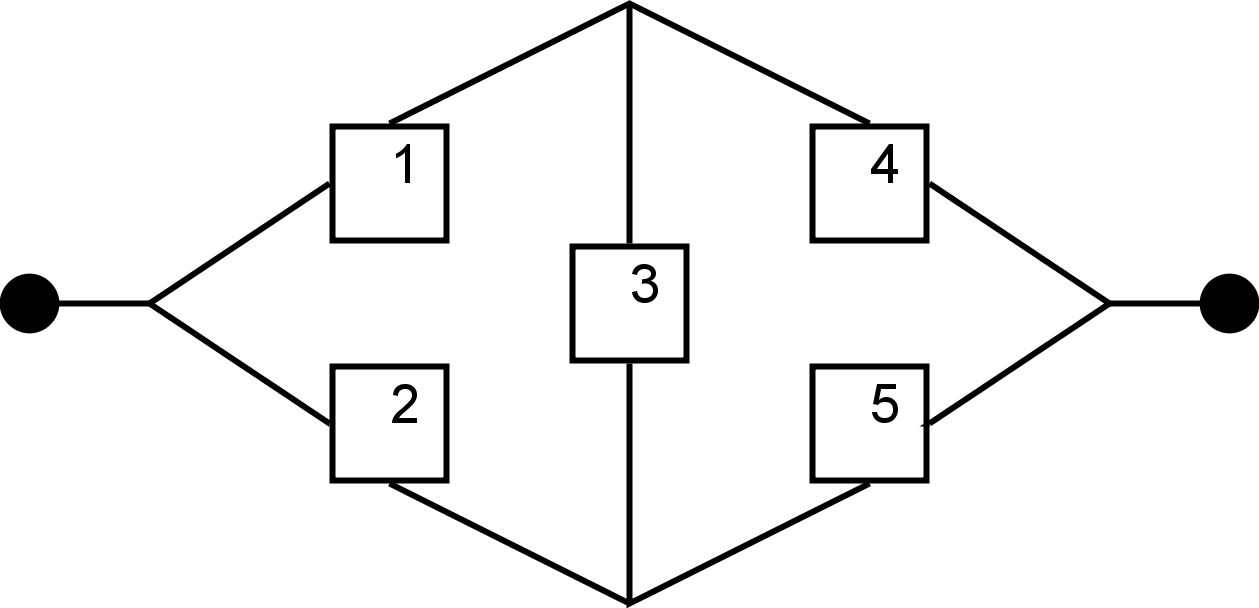}
	\caption{Bridge structure.}
	\label{fig:bridgenew}
\end{figure}

\begin{figure}[H]\centering
	\begin{minipage}[H]{0.3\linewidth}
		\includegraphics[width=\linewidth]{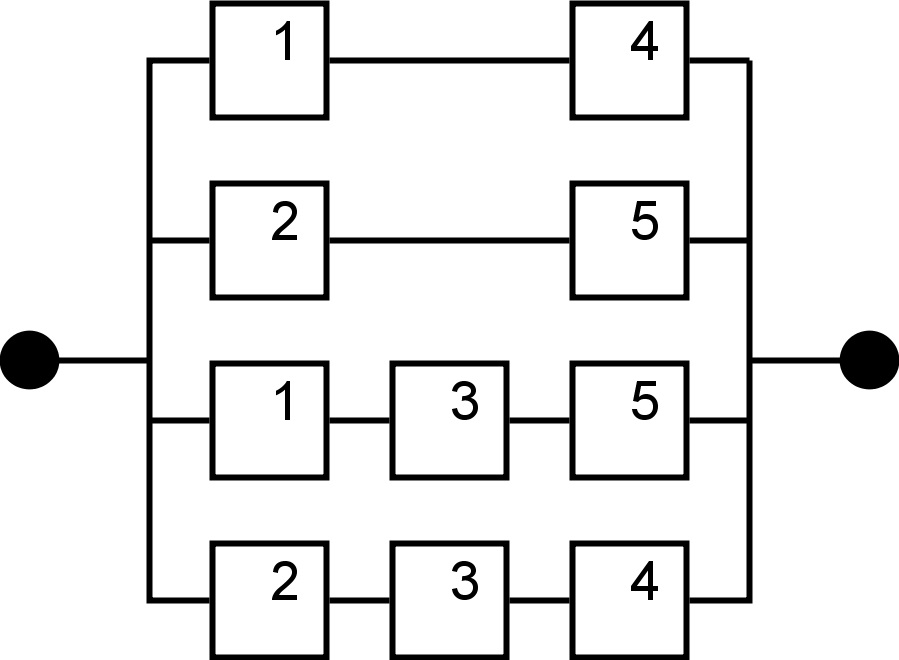}
		\subcaption{ $\mbox{   }$ } \label{bridge_PSS}
	\end{minipage}
	\begin{minipage}[H]{0.45\linewidth}
		\includegraphics[width=\linewidth]{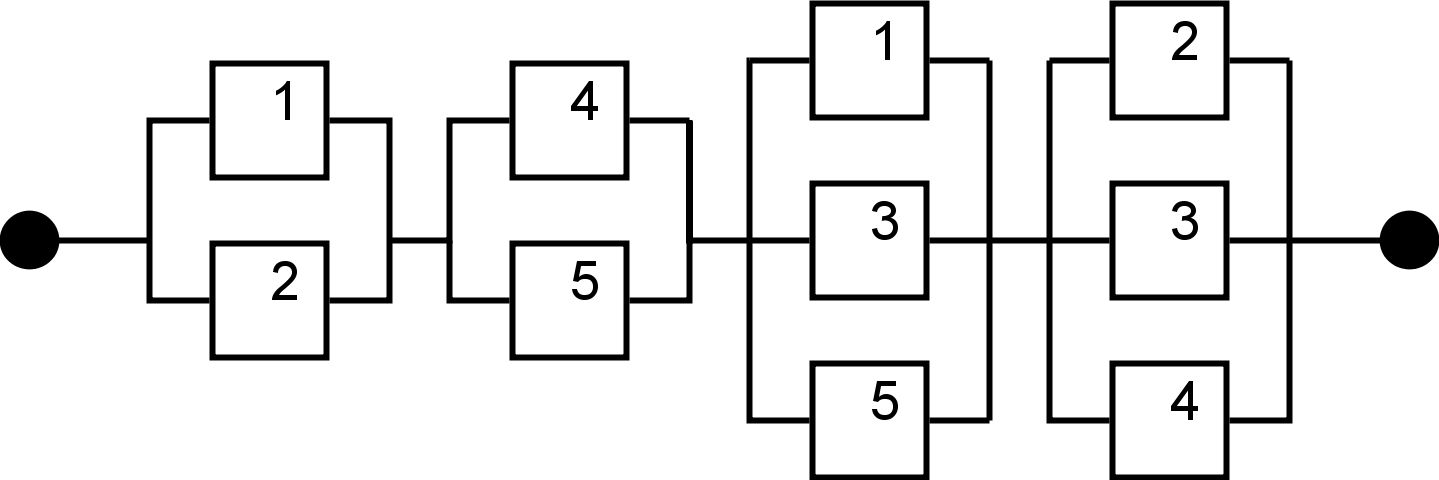}
		\subcaption{ $\mbox{   }$ } \label{bridge_SPS}
	\end{minipage}  
	\caption{(a) Bridge parallel-series representation. (b) Bridge series-parallel representation.}
	\label{system_bridge_SPS_PSS}
\end{figure}

\begin{figure}[H]\centering
	\begin{minipage}[H]{0.3\linewidth}
		\includegraphics[width=\linewidth]{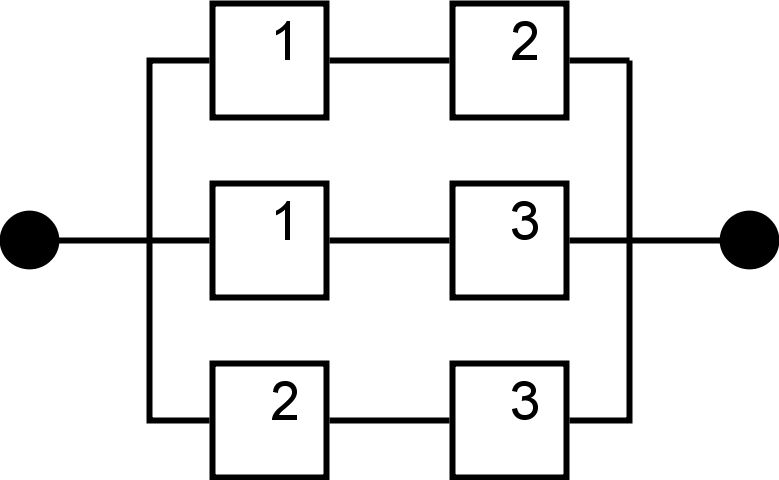}
		\subcaption{ $\mbox{   }$ } \label{2out3_PSS}
	\end{minipage}
	\begin{minipage}[H]{0.35\linewidth}
	\includegraphics[width=\linewidth]{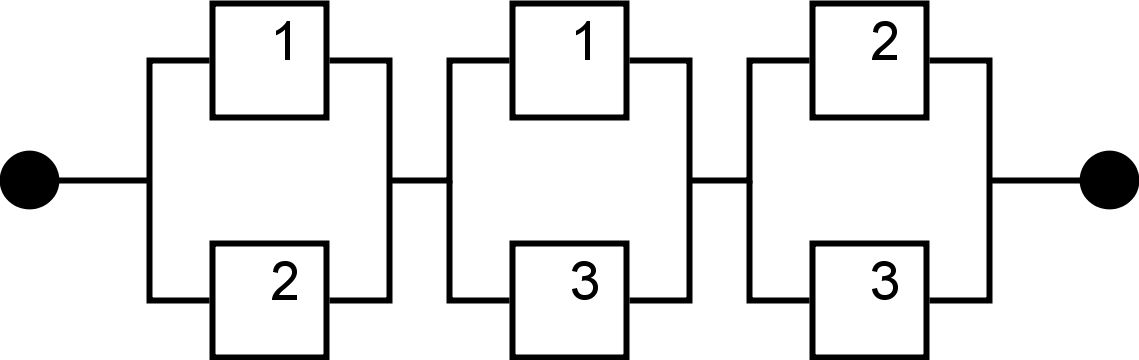}
	\subcaption{ $\mbox{   }$ }  \label{2out3_SPS}
\end{minipage} 
	\caption{(a) $2$-out-of-$3$ parallel-series representation (b) $2$-out-of-$3$ series-parallel representation.}
	\label{system_2de3_SPS_PSS}
\end{figure}

In general, for a coherent system that uses a representation combining parallel and serial systems, some components may appear in two or more places. Figure \ref{fig:bridgenew} is the bridge system described in the literature \citep{BProschan} and  Figure \ref{system_bridge_SPS_PSS} illustrates its parallel-series and series-parallel combinations. Note that each of the five  components appears twice for both representations. Another interesting structure is the $k$-out-of-$m$ system - it works only if at least $k$ out of the $m$ components work -. For instance, Figure \ref{system_2de3_SPS_PSS} considers the simple $2$-out-of-$3$ case into parallel-series and series-parallel representations. Note that each of the three components also appears twice in both combinations.  

\citet{Sassa} in their nonparametric inferences for coherent systems restricted themselves to the cases of independent and identically distributed failure times, hence all components have the same reliability.
The method introduced here does not need the supposition of independence and identically distributed component lifetimes. Anther restriction, considered and invisible in other methods in the literature, is that the failures of any pair of components cannot occur at the same time. This restriction is also not necessary here. Our main assumption is that all component lifetime distributions are the three-parameter Weibull, a very general distribution that can approximate most lifetime distributions. Besides the left- and right-censored observations, the interval-censored observations can also be handled here.  The paradigm of the present article is the Bayesian one. Another advantage of the Weibull is that, in our paradigm, even with improper priors, the posterior distributions turn out to be proper. The proposed mechanism of calculation introduced here can well be used for any other family of distributions whenever proper priors are used.

Section 2 describes the Bayesian Weibull model.  Section 3 presents the simulation studies used in most of our examples.  Section 3 compares the performance of the \citet{Sassa}  approach to ours. Section 4 illustrates the methodology by considering a practical example for which interval censoring appears. Final considerations appear in Section 5. The Appendix is devoted to showing that the posterior distributions are proper.

\newpage
	\section{A Weibull Reliability Model}\label{secao2}
	
Consider a system of $m$ components, $X_j$ denoting the failure time of the $j$-th component, $j = 1, \ldots , m$. A simple random sample of $n$ systems with the same structure is observed, $t_1, \ldots ,$ and $t_n$ being a sample of the random variable $T$, the system failure time.

The goal is to estimate the reliability of the system and for that the reliability of its components shall be estimated; the object of this paper. At system failure, however, not all components would have their failure time observed. In addition, a particular component may be responsible for system failures in some sample units and not in the remaining ones, cases of censoring on component failure time. The amount and the types of censoring depend on  the structure of the system. 
For example, a sample of $n$ units of a machine with configuration as in Figure \ref{PSS} is observed: The failure of component 1 alone is not enough for system failure. On the other hand, every failure of the system implies the failure of component 1, either being left-censored or being the last to fail. The latter should imply that one of the other two components, say component 2, had failure before component 1, a case of left-censoring.

 When a system fails, the failure time of a given component $j$ may not be observed, but its censored time of failure is.  For all sample units, the system failure times $t_1, \ldots ,$ and $t_n$ are recorded. For a specific component $j$, not responsible for one of the $n$ systems that failed at time $t$, either it is right-censored, in which case it still could continue to work after $t$, or it is censored to the left if it has failed before $t$.  Another kind of censoring could also occur:  Suppose a machine failure time (a sample unit) is in an interval $(L,U)$ – $L$ for the observed lower limit and $U$ for the upper limit.  If two or more components failed, they are all interval-censored in $(L,U)$. To generalize the notation for all cases of component failure and censoring, consider the following notation: for a specific component $j$ of system unit $i$, let $(L_{ji}, U_{ji})$ be a general interval of time, in which

	\begin{itemize}
		\item $L_{ji}=U_{ji}=t_i$, if the $j$-th component failure time causes the $i$-th system failure time;
		\item $L_{ji}=t_i$ and $U_{ji}=\infty$, if the $j$-th component is right-censored at $t_i$;
		\item $L_{ji}=0$ and $U_{ji}=t_i$ , if the $j$-th component is left-censored at $t_i$;
		\item $0 < L_{ji} < U_{ji} < \infty$, if the $j$-th component is interval-censored.
	\end{itemize} 

To complete the theoretical environment, let $X_j$ be the random variable representing the $j$-th component failure time with density function $f(x_j|\bm{\theta_j})$ and with reliability function  $R(x_j|\bm{\theta_j})$; $\bm{\theta_j}$ is the parameter, which can be either a scalar or a vector.

Using the above notation, the likelihood function is as follow:
   	\begin{eqnarray}
	{\rm L}(\bm{\theta_j} \mid {\bf l_{j}},{\bf u_{j}}) = \prod_{i=1}^n  {\Big[f(l_{ji}|\bm{\theta_j})\Big]}^{{\rm I}_{\{l_{ji}=u_{ji}\}}} {\Big[R(l_{ji}|\bm{\theta_j})
	-R(u_{ji}|\bm{\theta_j})\Big]}^{1-{\rm I}_{\{l_{ji}=u_{ji}\}}} , \label{veros}
	\end{eqnarray}
	where ${\rm I}_{\{TRUE\}}=1$ or ${\rm I}_{\{FALSE\}}=0$; ${\bf l_{j}}=(l_{j1},\ldots,l_{jn})$ and ${\bf u_{j}}=(u_{j1},\ldots,u_{jn})$.

	The likelihood function in (\ref{veros}) is generic and straightforward for any probability distribution. The distribution considered here is the three-parameter Weibull. The choice of this distribution is due to the variation of parameter values implying changes in both distribution shape and hazard rates. We can have increasing, decreasing and constant failure rates in this family of Weibull distributions \citep{Rinne}. 
	
	The Weibull reliability function is as follows:
	\begin{eqnarray}
	R(x_j \mid \bm{\theta_j}) = \exp\left[-\left(\frac{x_j-\mu_j}{\eta_j}\right)^{\beta_j}\right], \nonumber
	\end{eqnarray}
	for $x_j > 0$, where $\bm{\theta_j}= (\beta_j, \eta_j, \mu_j)$ and $\beta_j > 0$ (shape), $\eta_j > 0$ (scale) and $0<\mu_j<x_j$ (location). 
	
	The Weibull distribution with two parameters ($\mu_j=0$) is the most celebrated case in the literature.  However, the location parameter $\mu_j$ that represents the baseline lifetime has an important meaning in reliability and survival analysis. In reliability, a component under test may not be new. In medicine for instance, a patient may have the disease before the onset medical appoitment. Not taking account this initial time can underestimates the other parameters. Clearly, for new component testing $\mu_j$ may be $0$. 
	
	The posterior density of $\bm{\theta_j}= (\beta_j, \eta_j, \mu_j)$ comes out to be
	\begin{eqnarray}
	\pi(\beta_j, \eta_j, \mu_j \mid {\bf l_{j}},{\bf u_{j}}) \propto \pi(\beta_j, \eta_j, \mu_j) \prod_{i=1}^n   \left\{  \left( \frac{l_{ji}-\mu_j}{\eta_j} \right)^{\beta_j-1}\frac{\beta_j}{\eta_j} \exp \left[-\left(\frac{l_{ji}-\mu_j}{\eta_j} \right)^{\beta_j} \right] \right\}^{{\rm I}_{\{l_{ji}=u_{ji}\}}} \nonumber \\
 \times  \left\{ \exp \left[-\left(\frac{l_{ji}-\mu_j}{\eta_j}\right)^{\beta_j}\right] - \exp\left[-\left(\frac{u_{ji}-\mu_j}{\eta_j}\right)^{\beta_j}\right]\right\}^{1-{\rm I}_{\{l_{ji}=u_{ji}\}}} \label{posteriori}
	\end{eqnarray}
	for which $\pi(\beta_j, \eta_j, \mu_j)$ is the prior density of $\bm{\theta_j}=(\beta_j, \eta_j, \mu_j)$ that we consider to be:
	\begin{equation}
      \pi(\beta_j, \eta_j, \mu_j) =\frac{1}{\eta_j}\frac{1}{\beta_j}. \label{priori}
    \end{equation}	
	
%
%

     Even though (\ref{priori}) is not a proper prior – its integral is not finite – the posterior density in Equation (\ref{posteriori}) is still a proper, as stated by the following result.  
     
   	\begin{theorem} \label{theo}
   	Let a class of non-informative prior given by
   	\begin{eqnarray}
   	\pi(\beta_j, \eta_j, \mu_j)=\frac{1}{\eta_j \beta_j^b},~~ b\geq 0. \nonumber
   	\end{eqnarray}
   	 Although, for $b \geq 0$, $n = 1$ and the existence of a failure, the posterior in (\ref{posteriori}) is not proper, for $n > 1$, the posterior in (\ref{posteriori}) is proper. 
   \end{theorem}

   The proof of this result is left to the Appendix.

   	The importance of the above result is that one can perform Bayesian inferences even with little prior information. 
	
	
	Because the posterior density (\ref{posteriori}) does not have a closed form, statistical inferences about the parameters can rely on Markov-Chain Monte-Carlo (MCMC) simulations. Here we consider an adaptive Metropolis-Hasting algorithm with a multivariate distribution \citep{Haario}.
		
	Discarding burn-in (first generated values discarded to eliminate the effect of the assigned initial values for parameters) and jump samples (spacing among generated values to avoid correlation problems) a sample of size $n_p$ from the joint posterior distribution of $\bm{\theta_j}$ is obtained. For the $j$-th component, the sample from the posterior can be expressed as $(\beta_{j1},\beta_{j2},\ldots,\beta_{jn_p})$, $(\eta_{j1},\eta_{j2},\ldots,\eta_{jn_p})$ and $(\mu_{j1},\mu_{j2},\ldots,\mu_{jn_p})$. Consequently, posterior quantities of reliability function $R(t\mid \bm{\theta_j})$ can be easily obtained \citep{RobertCasella}. For instance, the posterior mean of the reliability function is
	\begin{eqnarray}
	{\rm E}[R(x_j\mid \bm{\theta_j}) \mid {\bf l_{j}},{\bf u_{j}}] = \frac{1}{n_p}\sum_{k=1}^{n_p}{R(x_j \mid \mbox{\boldmath{$\theta_{jk}$}})},~~ \mbox{for each} ~ t > 0. \label{relia_bayes}
	\end{eqnarray}
	\section{Model Evaluation with Simulated Data}
	\label{secao4}

	To evaluate the properness of the model described above, this section presents simulation studies in six scenarios with different generators of component lifetimes; also, different percentages of censored data are considered.  Two types of system structures are used: bridge system (Figure \ref{fig:bridgenew}) and $2$-out-of-$3$ system (Figure \ref{system_2de3_SPS_PSS}). For each scenario, five different samples of system units are considered ($n = 25$, $50$, $100$, $300$, $1000$). The observed kinds of information are the failure time of the $n$ observed systems and for each unit that failed the status of each component of that unit at the time of its failure.
	
	To obtain posterior quantities, we generated $20,000$ samples from the posterior distribution of each parameter. The first $10,000$ of these samples were discarded as burn-in samples. A jump of size $10$ was chosen to avoid correlation problems and, consequently, samples of size $n_p = 1,000$ were obtained for all estimation steps. The chains' convergence was monitored in all simulation scenarios for good convergence results to be obtained. The posterior mean W3PM (Weilbull three parameter means) is considered as the performance measure of the posterior reliability function obtained through our Weibull models. 
	
	The W3PM estimates are compared to the BSNP estimates, the nonparametric estimates of \cite{Sassa}. Their approach can be used for all components involved in the reliability of any system, even for more complex structures; the only necessary types of information for the computation of the estimates are the system structure and the observed system failure times. However, in their work, there is the strong restriction that all the component lifetimes are mutually independent and identically distributed. Consequently, all components have the same reliability.  The present approach does not have this limitation.  
	
    For each scenario, $1000$ copies (data sets) are generated and we evaluated the mean absolute error (MAE) from the estimators to the true distribution as the comparison measure.  $R(t)$ and $\widehat{R}(t)$ are the true reliability function and its estimate, respectively. Hence the MAE is evaluated by $\frac{1}{l}\sum_{\ell=1}^{l} \mid \widehat{R}(g_{\ell})-R(g_{\ell}) \mid$, where $\{g_1, \ldots, g_{\ell}, \ldots, g_l \}$ is a grid in the space of failure times.
	
	In what follows, six scenarios are presented:
   	\begin{itemize}
		\item {\bf Scenario 1}: $2$-out-of-$3$ structure in which $X_1$ was generated from a Weibull distribution with mean $15$ and variance $8$, $X_2$ from a gamma distribution with mean $18$ and variance $12$ and $X_3$ from a log-normal distribution with mean $20$ and variance $10$.
			
		\item {\bf Scenario 2}: $2$-out-of-$3$ structure in which $X_1$ was generated from a log-normal distribution with mean $4$ and variance $7$, $X_2$ from a modified Weibull distribution \citep{WeibMod} with mean $2.88$ and variance $12.44$ and $X_3$ was generated from three-parameter Weibull distribution with mean $5$ and variance $3$.

		\item {\bf Scenario 3}: $2$-out-of-$3$ structure in which $X_1$, $X_2$ and $X_3$ were generated from Weibull distributions with means $10$, $11$, $10$ and variances $2$, $10$, $8$, respectively.
		
		
		\item {\bf Scenario 4}: $2$-out-of-$3$ structure in which $X_1$, $X_2$ and $X_3$ were generated from modified Weibull distributions \citep{WeibMod} with means $1.6$, $2.4$, $2.9$ and variances $6$, $4$, $13$, respectively.

		\item {\bf Scenario 5}: Bridge structure in which $X_1$ was generated from a Weibull distribution with mean $17$ and variance $8$, $X_2$ from a log-normal distribution with mean $16$ and variance $22$, $X_3$ from a log-normal distribution with mean $15$ and variance $15$, $X_4$ from a gamma distribution with mean $15$ and variance $6$ and $X_5$ from a gamma distribution with mean $20$ and variance $12$.
		 
		\item {\bf Scenario 6}: Bridge tructure in which $X_1$ was generated from a Weibull distribution with mean $4$ and variance $15$, $X_2$ from a modified Weibull distribution \citep{WeibMod} with mean $5.6$ and variance $15$, $X_3$ from a log-normal distribution with mean $6$ and variance $7$, $X_4$ from a gamma distribution with mean $5$ and variance $8$ and $X_5$ from a three-parameter Weibull distribution with mean $4$ and variance $8$.		
	\end{itemize}
	
	Since a component that causes system failure leads the other components to be right- or left-censored data, the high percentages of censored data for all scenarios are shown in Table \ref{censored_data}. It can be noted that all components in all scenarios have high percentages of censored data, all cases higher than $50\%$, reaching up to $90\%$ (see component 3 in scenario 5).
	
\begin{figure}[h!]\centering
	\begin{minipage}[b]{0.48\linewidth}
		\includegraphics[width=\linewidth]{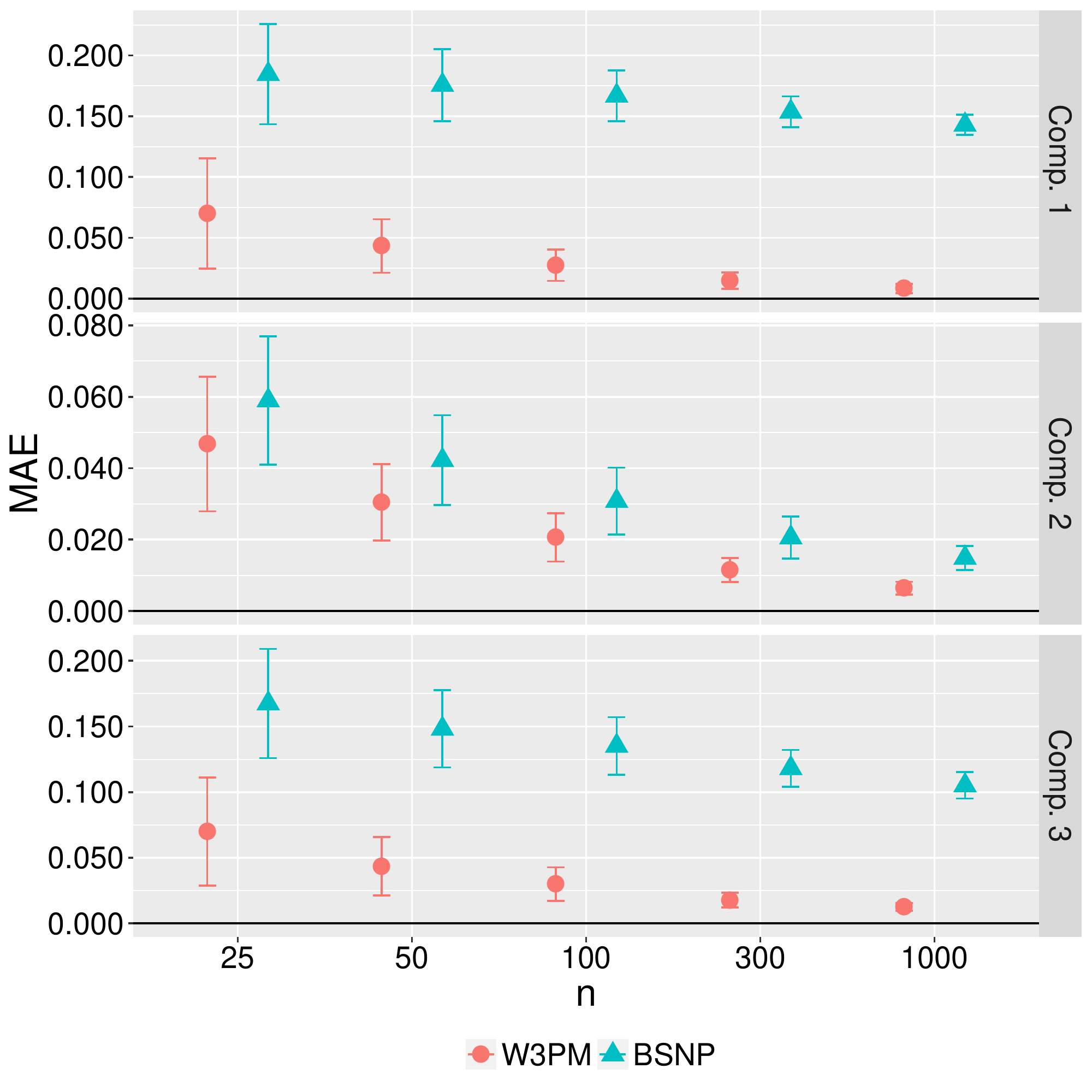}
		\subcaption{Scenario 1} \label{scenario1}
	\end{minipage} 
	\begin{minipage}[b]{0.48\linewidth}
		\includegraphics[width=\linewidth]{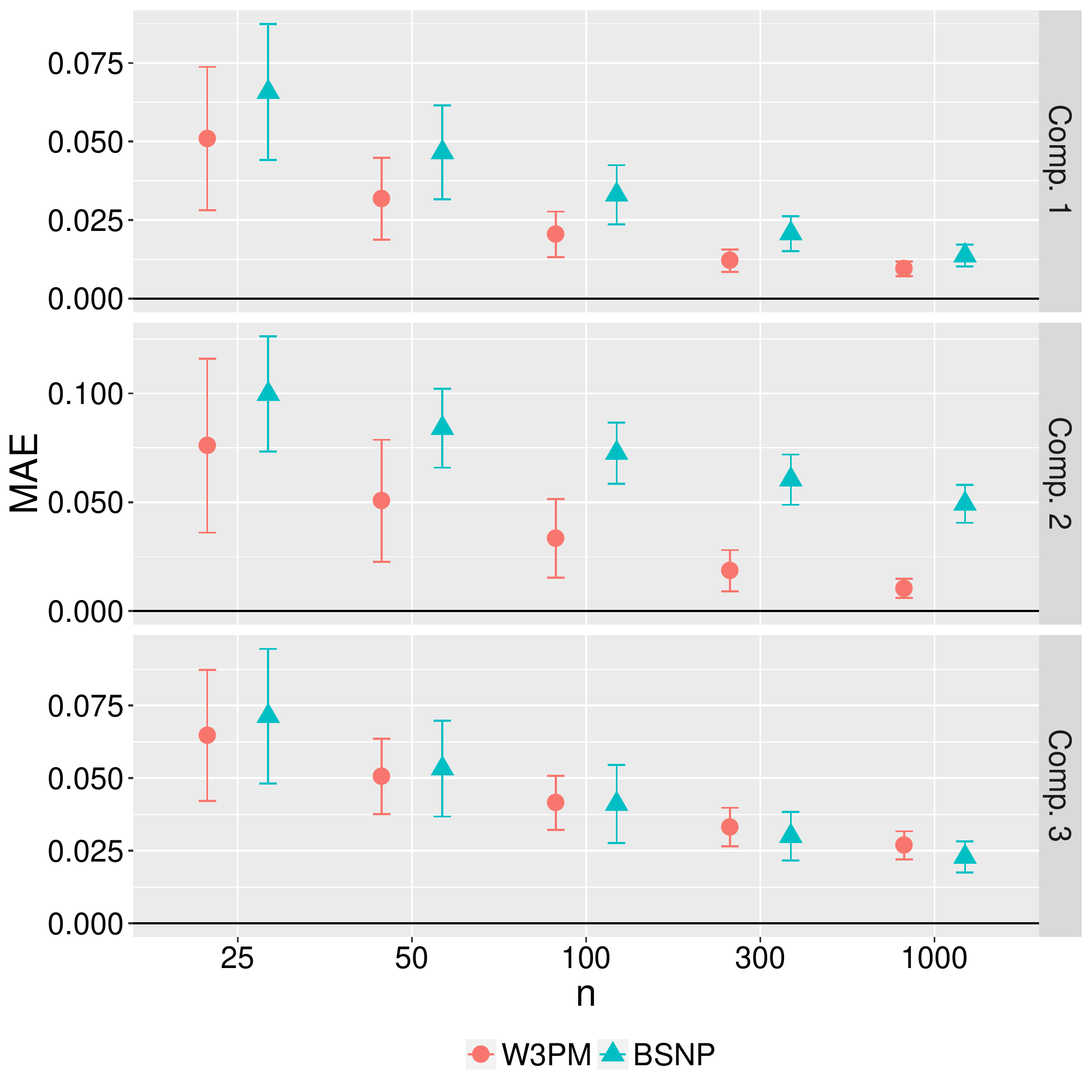}
		\subcaption{Scenario 2 } \label{scenario2}
	\end{minipage}
	\caption{Mean (symbol) and standard deviation (bars) of MAE obtined by W3PM and BSNP for scenarios 1 and 2.}
\end{figure}

\begin{figure}[h!]\centering
	\begin{minipage}[b]{0.48\linewidth}
		\includegraphics[width=\linewidth]{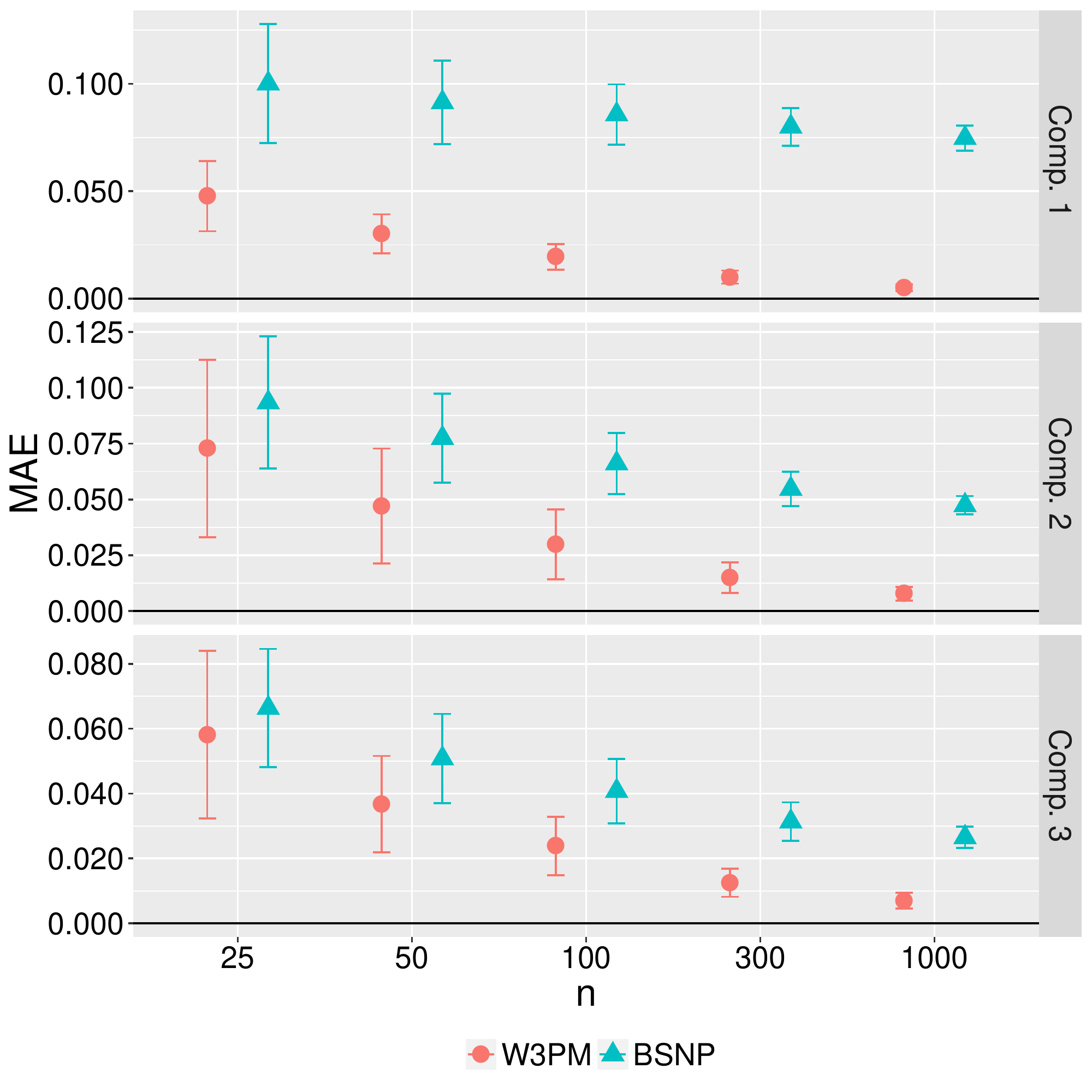}
		\subcaption{Scenario 3} \label{scenario3}
	\end{minipage} 
	\begin{minipage}[b]{0.48\linewidth}
		\includegraphics[width=\linewidth]{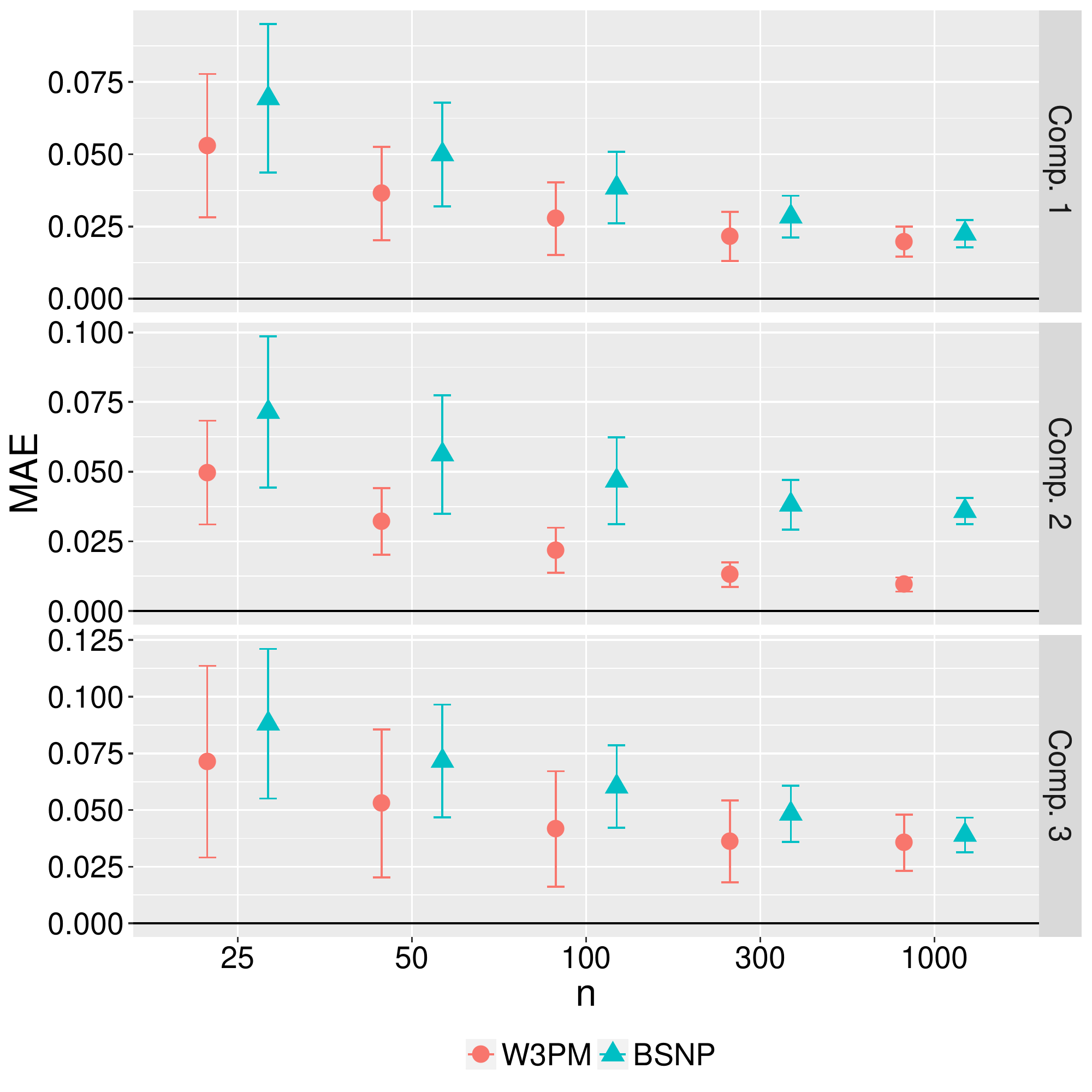}
		\subcaption{Scenario 4} \label{scenario4}
	\end{minipage}
	\caption{Mean (symbol) and standard deviation (bars) of MAE obtined by W3PM and BSNP for scenarios 3 and 4.}
\end{figure}

\begin{figure}[h!]\centering
	\begin{minipage}[b]{0.48\linewidth}
		\includegraphics[width=\linewidth]{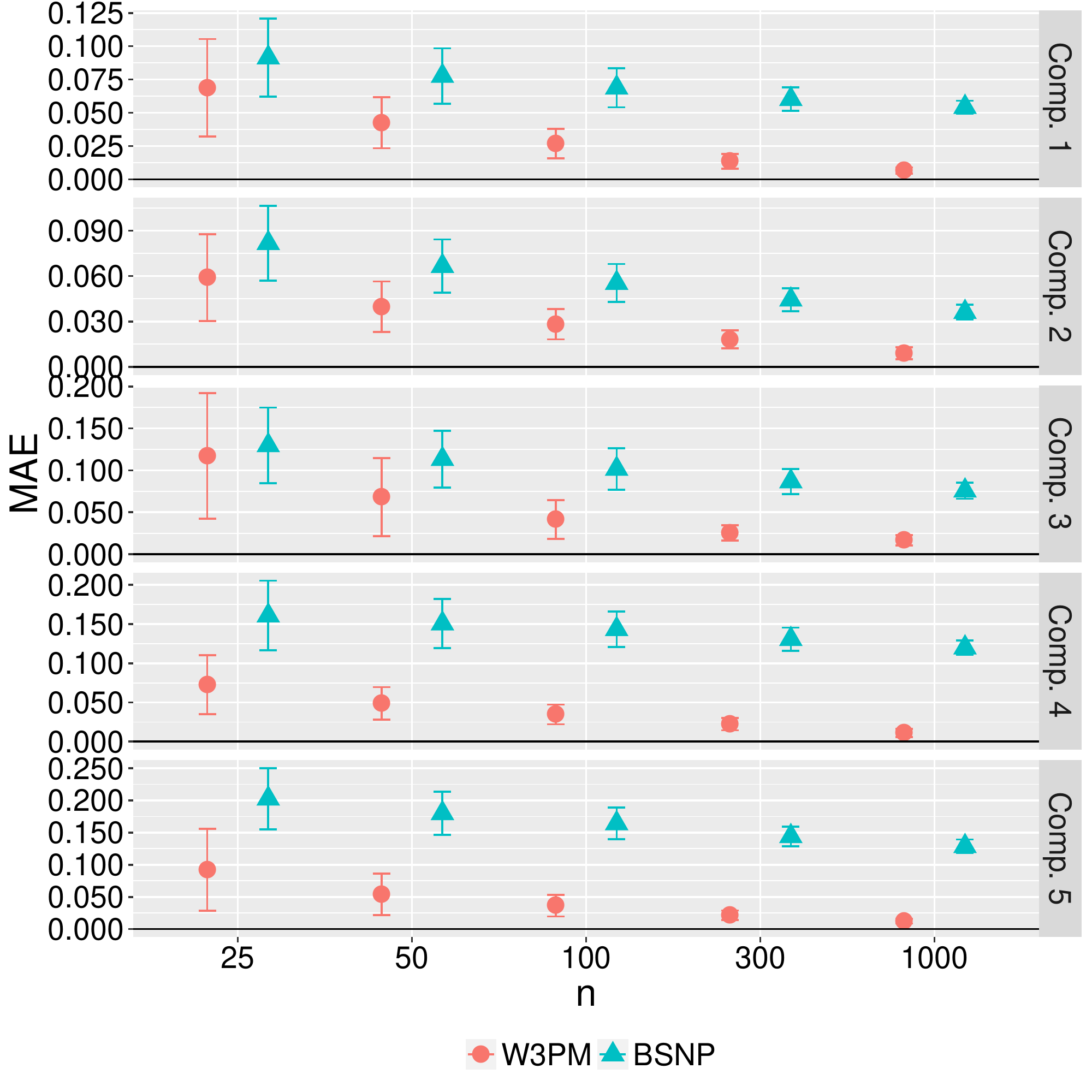}
		\subcaption{Scenario 5} \label{scenario5}
	\end{minipage} 
	\begin{minipage}[b]{0.48\linewidth}
		\includegraphics[width=\linewidth]{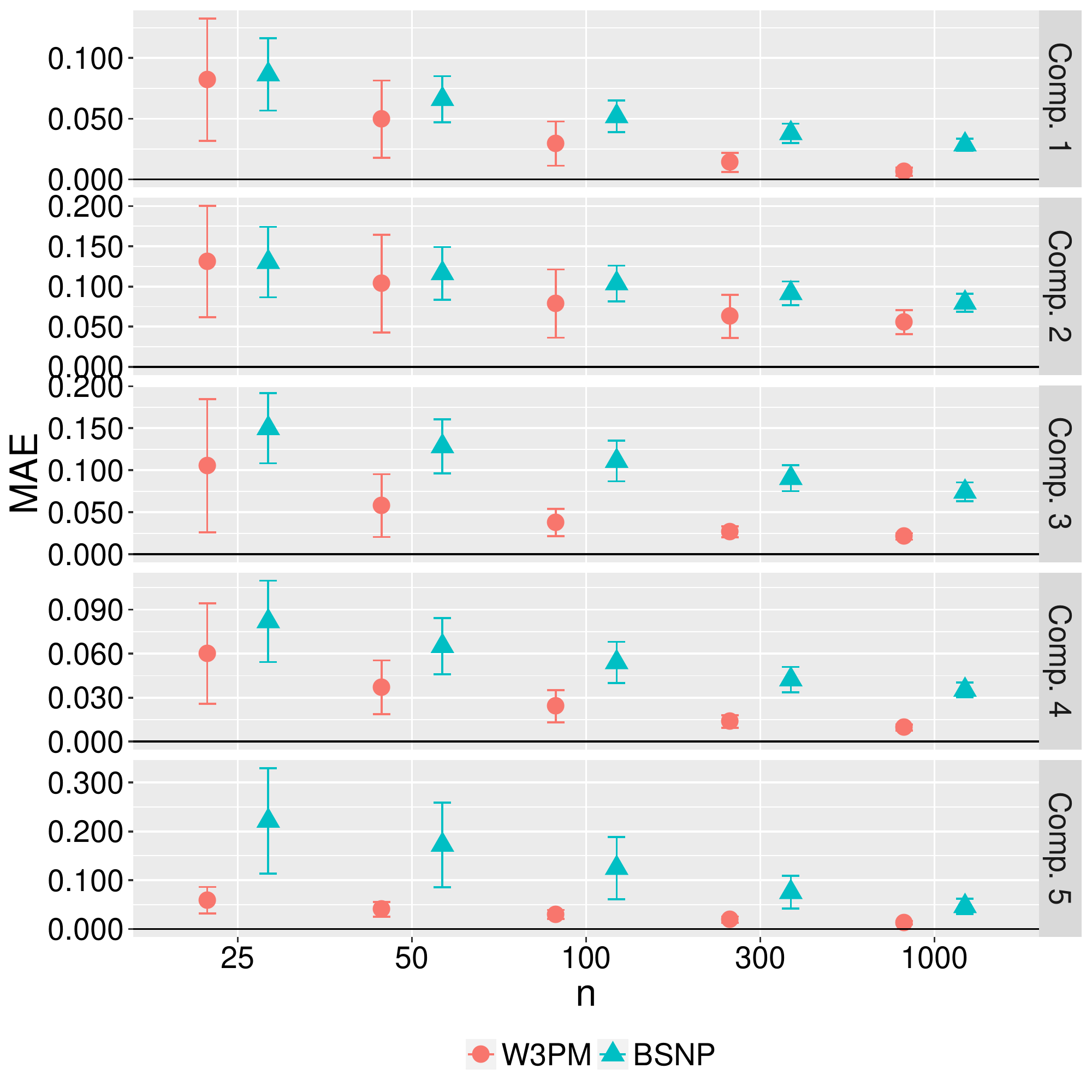}
		\subcaption{Scenario 6} \label{scenario6}
	\end{minipage}
	\caption{Mean (symbol) and standard deviation (bars) of MAE obtined by W3PM and BSNP for scenarios 5 and 6.}
\end{figure}

	The mean and stardard deviation of $1,000$ MAE values obtained by W3PM and BSNP are presented in Figures \ref{scenario1} to \ref{scenario6} for scenarios 1 to 6. For component 3 from scenarios 2 and 4 the two estimation methods showed similar behavior. For the other situations, W3PM always presents lower mean MAE values and the performance of the proposed estimator improves as $n$ increases.
	
\begin{table}[h!]
	\centering
	\scriptsize
	\caption{Percentage of censored data for each component in each scenario.}
	\begin{tabular}{ccccc}
		\hline
		Scenario & Component & \multicolumn{3}{c}{Side of censoring} \\
		&       & Left  & Right & Total \\
		\hline
		& 1 & 72.40\% & 5.40\% & 77.80\% \\
		1	& 2 & 21.10\% & 33.30\% & 54.40\% \\
		&   3 & 6.50\% & 61.30\% & 67.80\% \\
		\hline
		&  1   & 37.40\% & 25.70\% & 63.10\% \\
		2 &   2 & 48.00\% & 32.80\% & 80.80\% \\
		&   3 & 14.60\% & 41.50\% & 56.10\% \\
		\hline
		&  1   & 34.20\% & 21.30\% & 55.50\% \\
		3	&   2 & 27.20\% & 49.10\% & 76.30\% \\
		&  3 & 38.60\% & 29.60\% & 68.20\% \\
		\hline
		&  1   & 40.00\% & 30.50\% & 70.50\% \\
		4 &  2 & 32.40\% & 25.60\% & 58.00\% \\
		&   3 & 27.60\% & 43.90\% & 71.50\% \\
		\hline
		&  1   & 28.70\% & 51.40\% & 80.10\% \\
		&   2 & 50.40\% & 18.90\% & 69.30\% \\	
		5 &   3 & 62.50\% & 27.50\% & 90.00\% \\
			&  4 & 61.70\% & 19.10\% & 80.80\% \\
		&   5 & 8.50\% & 71.30\% & 79.80\% \\
		\hline
		&  1   & 52.60\% & 33.30\% & 85.90\% \\
		& 2 & 24.30\% & 59.50\% & 83.80\% \\
		6 & 3 & 17.90\% & 69.50\% & 87.40\% \\
		  &  4 & 23.00\% & 39.70\% & 62.70\% \\
		& 5 & 64.20\% & 16.00\% & 80.20\% \\
		\hline
	\end{tabular}%
	\label{censored_data}%
\end{table}

\clearpage

\section{Application}\label{aplication}

To show the applicability of the proposed method, a social study is considered in which the proposed methodology can be suitable applied. The data is analyzed by \citet{Klein}. In this study $n=191$ California high school boys were asked about their first use marijuana. The answers were either age in years, if the responder did use and remember the age, or ``I never used it'', which are right-censored observations at the boys’ current ages, or ``I have used it but I cannot remember the exactly time for my first use of the drug''. The latter is a left-censored observation case \citep{Klein}.

Klein and Moeschberger analysed the data through Turnbull's estimator \citep{turnbull1976}. In their approach, boys who remember their ages at the first time they use the drug produced uncensored observations. Consider, for instance,
a boy saying that he used for the first time the drug at $13$ years old and specifically it happens when he was $13$ years and $11$ months old. He would be considered a subject with uncensored observation at the age of $13$ by the Klein and Moeschberger analysis, even if his age of use was closer to $14$ years.  We believe that this kind of information should be considered as an interval-censored observation at $[13;14)$ and his information would be properly taken as interval-censored in our likelihood, the second factor of the right-hand side of equation (\ref{veros}).  
In this way, all data are censored: either right-,  left-, or interval-censored data.

To obtain posterior quantities related to the posterior distribution of $\bm{\theta}=(\beta,\eta,\mu)$ from posterior distribution in (\ref{posteriori}) through MCMC simulations, we discarded the first $10,000$ as burn-in samples and jump of size $30$ to avoid correlation problems, obtaining a sample of size $n_p=1,000$. The chains' convergence was monitored for all simulation scenarios and good convergence results were obtained.

\begin{figure}
	\centering
	\includegraphics[width=0.5\linewidth]{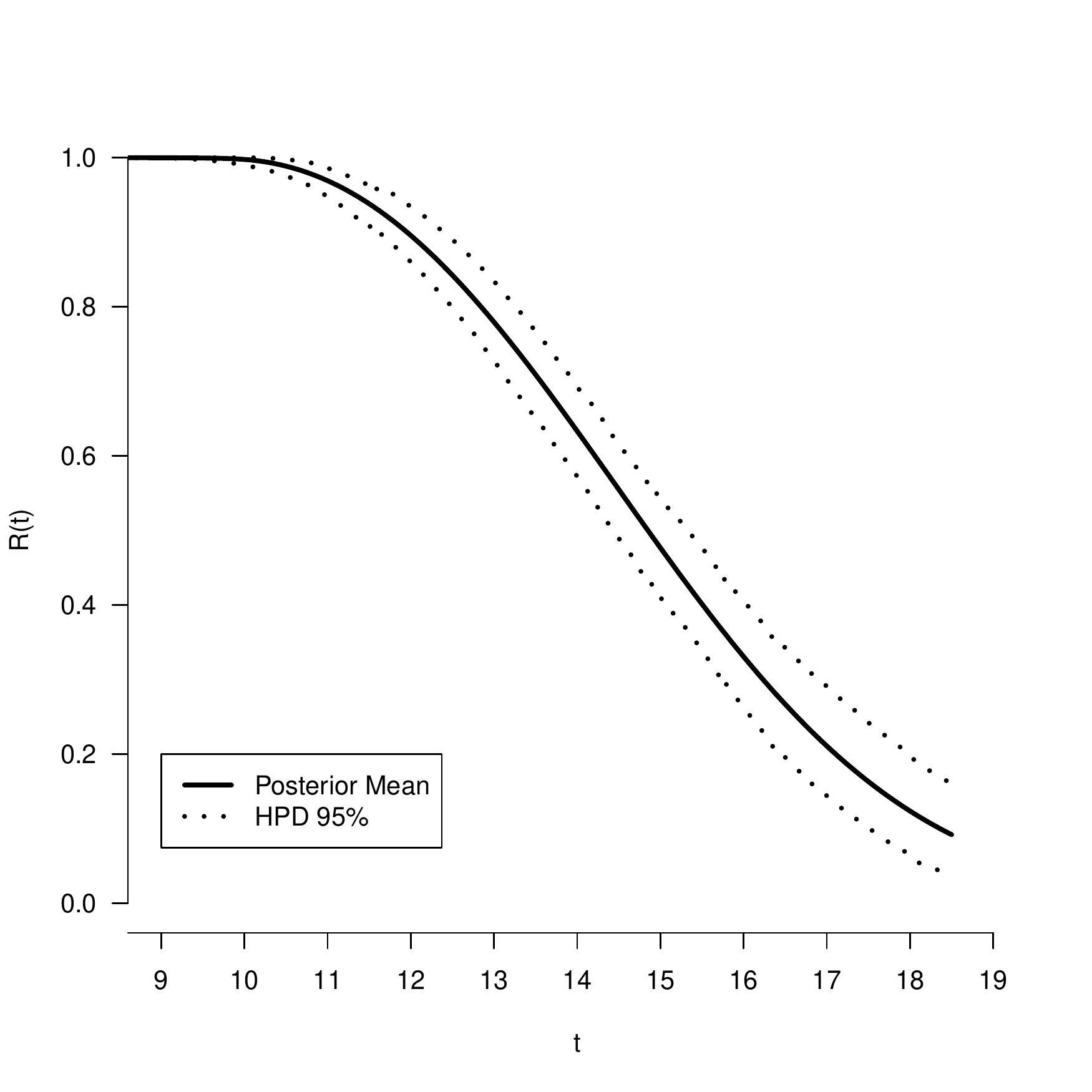}
	\caption{Posterior mean and 95\% HPD band of reliability function for time to first use of marijuana.}
	\label{fig:fumoconfia}
\end{figure}

Table \ref{fumo_posterior} lists the posterior means and posterior standard deviation for the parameters of shape ($\beta$), scale ($\eta$), location ($\mu$) and expected time of first use of marijuana - $\rm{E}(X|\bm{\theta})=\mu+\eta\Gamma(1+(1/\beta))$ -. The posterior mean of expected time of first use marijuana is $15.05$ years. The posterior mean and 95\% highest posterior density (HPD) band of the reliability function are illustrated by Figure \ref{fig:fumoconfia}. 

\begin{table}[htbp]
	\centering
	\caption{Posterior mean and posterior standard deviation of proposed model parameters and mean time of first use marijuana.}
	\begin{tabular}{ccc}
		\hline
		& Posterior mean & Posterior standard deviation \\
		\hline
		$\beta$  & 2.4   & 0.37 \\
		$\eta$   & 6.19  & 0.58 \\
		$\mu$    & 9.54  & 0.52 \\
		$\rm{E}(X|\bm{\theta})$ & 15.05 & 0.23 \\
		\hline
	\end{tabular}%
	\label{fumo_posterior}%
\end{table}%

\clearpage

\section{Final Remarks} \label{consideracoes}

A Bayesian Weibull model for component reliability was proposed. Neither independence nor identical distributions of component lifetimes was imposed. The proposed methodology is said to be general because it can be used for any coherent system, from the simplest to more complex structures.  Besides, it can also be appropriate for all kinds of censored data, including interval-censored, allowing it to be used in survival problems.  In estimation processes, satisfactory results about convergence were obtained and it was proved that the posterior is proper even when using prior distributions chosen from a family of non-informative prior distributions. We worked with the Bayesian Weibull model; however, it is quite simple to extend the work to other distributions or even to the pure likelihood approach \citep{PereiraPereira}.

\citet{Sassa} also considered complex coherent structures as we have here, however, the assumption of independent and identically distributed component lifetimes excludes the use of their method for most practical applications. On the other hand, their methodology does not require the choice of a parametric family of distributions. For positive random variables, we believe that the three-parameter Weibull family is a very rich family since most real situations will have random aspects that can be represented by an element of the family. 

Both methods were evaluated in scenarios with different distributions for the generation of component lifetimes, different percentages of censored data and different sample sizes. The observed information consists of the failure time of systems and the status of each component at the moment of each system failure. The simulation study showed excellent performance of the proposed estimator, and that its advantage increases with the sample size. For the cases where the W3PM did not perform better, it was in fact very close to the performance of BSNP.  
 
The practical relevance was assessed in a real dataset of boys' first use marijuana in which all observations are censored data and the proposed methodology can be suitably applied, showing that it can be used in other areas beyond reliability.
We also believe that for future work our methodology can be used to do reverse engineering.  Using model selection Bayesian techniques as in \citet{PereiraPereiraBook} that use mixture of reliability estimates one can choose one out of some alternative reliability systems.

\clearpage
\section*{Appendix} \label{apendice}

\subsection*{Proof of Theorem \ref{theo}}

	

		We have to show that 
		\begin{eqnarray}
		\int_{0}^{\min\{{\bf t}\}}\int_{0}^{\infty}\int_{0}^{\infty}\pi(\beta_j,\eta_j,\mu_j\mid  {\bf l_{j}},{\bf u_{j}})d\beta_j d\eta_j d\mu_j < \infty, \nonumber
		\end{eqnarray}
		where 
		\begin{eqnarray}
		\pi(\beta_j,\eta_j,\mu_j\mid  {\bf l_{j}},{\bf u_{j}}) \propto \pi(\beta_j,\eta_j,\mu_j)\prod_{i=1}^n\left\{  \left( \frac{l_{ji}-\mu_j}{\eta_j} \right)^{\beta_j-1}\frac{\beta_j}{\eta_j} \exp \left[-\left(\frac{l_{ji}-\mu_j}{\eta_j} \right)^{\beta_j} \right] \right\}^{{\rm I}_{\{l_{ji}=u_{ji}\}}} \nonumber \\
		\times  \left\{ \exp \left[-\left(\frac{l_{ji}-\mu_j}{\eta_j}\right)^{\beta_j}\right] - \exp\left[-\left(\frac{u_{ji}-\mu_j}{\eta_j}\right)^{\beta_j}\right]\right\}^{1-{\rm I}_{\{l_{ji}=u_{ji}\}}} . \nonumber
		\end{eqnarray}
		
		As this proof works for all $j$, we will omit the $j$ index. 
		
		For $n=1$ and $l_{1}=u_{1}$, 
		\begin{eqnarray}
		\int_{0}^{l_{1}}\int_{0}^{\infty} \int_{0}^{\infty}\frac{1}{\eta\beta^b}\left( \frac{l_{1}-\mu}{\eta} \right)^{\beta-1}\frac{\beta}{\eta} \exp \left[-\left(\frac{l_{1}-\mu}{\eta} \right)^{\beta} \right] d\beta d\eta d\mu &=&  \nonumber \\
		\int_{0}^{l_{1}}\int_{0}^{\infty}\frac{1}{\beta^b}\int_{0}^{\infty}\left[\frac{\beta (l_{1}-\mu)^{\beta-1}}{\eta^{\beta+1}}\exp\left\{-\left(\frac{l_{1}-\mu}{\eta}\right)^{\beta}\right\}\right] d\eta d\beta d\mu. \label{int_n1}
		\end{eqnarray}
		
		Let $X$ be a random variable that, given $\alpha$ and $\gamma$, follows and inverse gamma distribution and its density function is expressed as
		\begin{eqnarray}
		f(x\mid\alpha,\gamma)=\frac{\gamma^\alpha}{\Gamma(\alpha)}x^{-\alpha-1}\exp\left\{-\frac{\gamma}{x}\right\}, ~~\alpha >0 ~~ \mbox{and} ~~ \gamma >0. \nonumber 
		\end{eqnarray}
		Consider the variable change: $[(l_{1}-\mu)/\eta]^\beta=\gamma/x$, from which it follows that $(l_{1}-\mu)^{\beta}dx=\gamma\beta\eta^{\beta-1}d\eta$, so the integral expression in (\ref{int_n1}) can be written as
		\begin{eqnarray}
		\int_{0}^{l_{1}}\int_{0}^{\infty}\frac{1}{\beta^b}\int_{0}^{\infty}\left(\frac{\gamma}{x}\right)^2\frac{1}{(l_{1}-\mu)\gamma}\exp\left\{-\frac{\gamma}{x}\right\}dxd\beta d\mu &=& \int_{0}^{l_{1}}\int_{0}^{\infty}\frac{1}{\beta^b(l_{1}-\mu)}\int_{0}^{\infty}\gamma x^{-2}\exp\left\{-\frac{\gamma}{x}\right\}dxd\beta d\mu \nonumber \\
		&=& \int_{0}^{l_{1}} \int_{0}^{\infty}\frac{1}{\beta^b(l_{1}-\mu)}d\beta  d\mu \nonumber \\
		&=& \infty. \nonumber
		\end{eqnarray}
		In summary, for $n=1$ and $l_{1}=u_{1}$,
		\begin{eqnarray}
		\int_{0}^{l_{1}}\int_{0}^{\infty}\int_{0}^{\infty}\pi(\beta,\eta,\mu \mid l_{1},u_{1})d\beta d\eta d\mu=\infty. \nonumber 
		\end{eqnarray}
		
	Consider that for a sample of size $n$, $n> 1$, data are observed such that $l_{i}=u_{i}$, for $i=1,2,\ldots,n_f$ and $l_{i}\neq u_{i}$ for $i=n_{f+1},\ldots,n$.
\begin{eqnarray} 
\pi(\beta,\eta,\mu \mid {\bf l},{\bf u}) \propto \pi(\beta,\eta,\mu)\prod_{i=1}^{n_f}\left[\left(\frac{l_{i}-\mu}{\eta}\right)^{\beta-1}\frac{\beta}{\eta}\exp\left\{-\left(\frac{(l_{i}-\mu)}{\eta}\right)^{\beta}\right\}\right]  \nonumber \\
\times \prod_{i=n_f+1}^{n}\left[\exp\left\{-\left(\frac{l_{i}-\mu}{\eta}\right)^{\beta}\right\}-\exp\left\{-\left(\frac{u_{i}-\mu}{\eta}\right)^{\beta}\right\}\right],\nonumber
\end{eqnarray}
where ${\bf t_{l}}=(t_{l1},\ldots,t_{ln})$, for $l=1,2$.

As $l_{i}<u_{i}$, for all $i=1,\ldots,n$, we have that 
\begin{eqnarray}
\exp\left\{-\left(\frac{l_{i}-\mu}{\eta}\right)^{\beta}\right\} >  \exp\left\{-\left(\frac{u_{i}-\mu}{\eta}\right)^{\beta}\right\}. \nonumber
\end{eqnarray}

This way, 
\begin{eqnarray}
&&\pi(\beta,\eta,\mu \mid {\bf l},{\bf u}) \propto \nonumber \\   && ~~~~ \pi(\beta,\eta,\mu)\prod_{i=1}^{n_f}\left[\left(\frac{l_{i}-\mu}{\eta}\right)^{\beta-1}\frac{\beta}{\eta}\exp\left\{-\left(\frac{l_{i}-\mu}{\eta}\right)^{\beta}\right\}\right] \prod_{i=n_f+1}^{n}\left[\exp\left\{-\left(\frac{l_{i}-\mu}{\eta}\right)^{\beta}\right\}-\exp\left\{-\left(\frac{u_{i}-\mu}{\eta}\right)^{\beta}\right\}\right]  \nonumber \\
&& ~~~~ < \pi(\beta,\eta,\mu)\prod_{i=1}^{n_f}\left[\left(\frac{l_{i}-\mu}{\eta}\right)^{\beta-1}\frac{\beta}{\eta}\exp\left\{-\left(\frac{l_{i}-\mu}{\eta}\right)^{\beta}\right\}\right] \prod_{i=n_f+1}^{n}\left[\exp\left\{-\left(\frac{l_{i}-\mu}{\eta}\right)^{\beta}\right\}\right]. \nonumber
\end{eqnarray}
Thus, it is necessary only to evaluate the upper bond, that is, 
\begin{eqnarray}
\int_{0}^{\min\{{\bf t}\}}\int_{0}^{\infty} \int_{0}^{\infty}\pi(\beta,\eta,\mu)\prod_{i=1}^{n_f}\left[\left(\frac{l_{i}-\mu}{\eta}\right)^{\beta-1}\frac{\beta}{\eta}\exp\left\{-\left(\frac{l_{i}-\mu}{\eta}\right)^{\beta}\right\}\right] \prod_{i=n_f+1}^{n}\left[\exp\left\{-\left(\frac{l_{i}-\mu}{\eta}\right)^{\beta}\right\}\right]d\beta d\eta d\mu < \infty. \label{prova_posteriori}
\end{eqnarray} 

Let $t_{mi}=l_{i}-\mu$ and consider first the integrals in $\beta$ and $\eta$, that is, 
\begin{eqnarray}
{\rm I}&&=\int_{0}^{\infty} \int_{0}^{\infty}\pi(\beta,\eta,\mu)\prod_{i=1}^{n_f}\left[\left(\frac{t_{mi}}{\eta}\right)^{\beta-1}\frac{\beta}{\eta}\exp\left\{-\left(\frac{t_{mi}}{\eta}\right)^{\beta}\right\}\right] \prod_{i=n_f+1}^{n}\left[\exp\left\{-\left(\frac{t_{mi}}{\eta}\right)^{\beta}\right\}\right]d\beta d\eta \nonumber \\
&& =\int_{0}^{\infty} \int_{0}^{\infty}\frac{1}{\beta^b\eta}\prod_{i=1}^{n_f}\left[\left(\frac{t_{mi}}{\eta}\right)^{\beta-1}\frac{\beta}{\eta}\right] \prod_{i=1}^{n}\left[\exp\left\{-\left(\frac{t_{mi}}{\eta}\right)^{\beta}\right\}\right]d\beta d\eta  \nonumber \\
&& =\int_{0}^{\infty} \int_{0}^{\infty}\frac{1}{\beta^b\eta}\frac{\beta^{n_f}}{\eta^{\beta n_f}}\prod_{i=1}^{n_f}{\left[t_{mi}^{\beta-1}\right]} \exp\left\{-\sum_{i=1}^n\left(\frac{t_{mi}}{\eta}\right)^{\beta}\right\}d\beta d\eta. \label{prova_propria_n}
\end{eqnarray}

Consider again the variable change: $\left(\sum_{i=1}^n t_{mi}^{\beta}/\eta^{\beta}\right)=\gamma/x$, from which it follows that $\sum_{i=1}^n t_{mi}^{\beta}dx=\gamma\beta\eta^{\beta-1}d\eta$. So we can write the expression in (\ref{prova_propria_n}) as
\begin{eqnarray}
&& \int_{0}^{\infty}\frac{\beta^{-b}}{\left[\sum_{i=1}^n t_{mi}^{\beta}\right]^{n_f}}\prod_{i=1}^{n_f}{\left[t_{mi}^{\beta-1}\right]}\int_{0}^{\infty} \beta^{n_f-1}\gamma^{n_f}x^{-n_f-1}\exp\left\{-\frac{\gamma}{x}\right\} dx d\beta  \nonumber \\ 
&=&\int_{0}^{\infty}\frac{\beta^{n_f-b-1}}{\left[\sum_{i=1}^n t_{mi}^{\beta}\right]^{n_f}}\prod_{i=1}^{n_f}{\left[t_{mi}^{\beta-1}\right]}\Gamma(n_f)\int_{0}^{\infty} \frac{\gamma^{n_f}}{\Gamma(n_f)}x^{-n_f-1}\exp\left\{-\frac{\gamma}{x}\right\} dx d\beta  \nonumber \\
&=&\int_{0}^{\infty}\frac{\beta^{n_f-b-1}}{\left[\sum_{i=1}^n t_{mi}^{\beta}\right]^{n_f}}\prod_{i=1}^{n_f}{\left[t_{mi}^{\beta-1}\right]}\Gamma(n_f) d\beta.  \nonumber
\end{eqnarray}

Let $c$ be a real positive number such that $c > \prod_{i=1}^{n_f}{t_{mi}^{-1}}$ and consider also $t_{1v}$ such that $t_{1v} < \max{(t_{m1},\ldots,t_{mn_f})}$. This way, for all $b > 0$, we have that
\begin{eqnarray}
&& \int_{0}^{\infty}\beta^{n_f-b-1}\Gamma(n_f)\frac{\prod_{i=1}^{n_f}{t_{mi}^{\beta-1}}}{\left[\sum_{i=1}^n t_{mi}^{\beta}\right]^{n_f}} d\beta < \int_{0}^{\infty}c\beta^{n_f-b-1}\Gamma(n_f)\frac{\prod_{i=1}^{n_f}{t_{mi}^{\beta}}}{\left[\sum_{i=1}^n t_{mi}^{\beta}\right]^{n_f}} d\beta  \nonumber \\ 
&& < \int_{0}^{\infty}c\beta^{n_f-b-1}\Gamma(n_f)\frac{\prod_{i=1}^{n_f}{t_{mi}^{\beta}}}{\left[\sum_{i=1}^{n_f}t_{mi}^{\beta}\right]^{n_f}} d\beta < c\Gamma(n_f)\int_{0}^{\infty}\beta^{n_f-b-1}\left[\frac{t_{1v}}{\max(t_{m1},\ldots,t_{mn_f})}\right]^{\beta} d\beta.  \label{int_penultimo}
\end{eqnarray}

Let $h=t_{1v}/\max(t_{m1},\ldots,t_{mn_f})$. We can write the last expression in (\ref{int_penultimo}) as
\begin{eqnarray}
c\Gamma(n_f)\int_{0}^{\infty}\beta^{n_f-b-1}h^{\beta} d\beta=  c\Gamma(n_f)\int_{0}^{\infty}\beta^{n_f-b-1}\exp\left\{\beta\ln(h)\right\}d\beta.\label{int_h}
\end{eqnarray}

Considering the variable change $v=-\beta\ln(h)$, we have that $dv=-\ln(h)d\beta$ and
(\ref{int_h}) can be expressed as
\begin{eqnarray}
c\Gamma(n_f)\int_{0}^{\infty}\Bigg(\frac{1}{|\ln(h)|}\Bigg)^{n_f-b}v^{n_f-b-1}\exp\{-v\}dv, \nonumber
\end{eqnarray}

since $h<1$. Let $c_2= c\Gamma(n_f)\Bigg(\frac{1}{|\ln(h)|}\Bigg)^{n_f-b}$. This way, 
\begin{eqnarray}
c_2\int_{0}^{\infty}v^{n_f-b-1}\exp\{-v\}dt < c_2 \int_{0}^{\infty}v^{z-1}\exp\{-v\}dv < \infty, \label{int_last}
\end{eqnarray}
where $z$ is the smallest positive integer bigger than $n_f-b$.   

The result in (\ref{int_last}) is valid since, for all positive integer $a$, 
\begin{eqnarray}
\Gamma(a)=\int_{0}^\infty v^{a-1}\exp\{-v\}dv <  \infty. \nonumber
\end{eqnarray}

This way, we have that
\begin{eqnarray}
{\rm I}&&=\int_{0}^{\infty} \int_{0}^{\infty}\pi(\beta,\eta,\mu)\prod_{i=1}^{n_f}\left[\left(\frac{t_{mi}}{\eta}\right)^{\beta-1}\frac{\beta}{\eta}\exp\left\{-\left(\frac{t_{mi}}{\eta}\right)^{\beta}\right\}\right] \prod_{i=n_f+1}^{n}\left[\exp\left\{-\left(\frac{t_{mi}}{\eta}\right)^{\beta}\right\}\right]d\beta d\eta < \infty \label{finito}
\end{eqnarray}

Returning to equation (\ref{prova_posteriori}) and considering (\ref{finito}), we finally have that  
\begin{eqnarray}
\int_{0}^{\min\{{\bf t}\}}\int_{0}^{\infty} \int_{0}^{\infty}\pi(\beta,\eta,\mu)\prod_{i=1}^{n_f}\left[\left(\frac{t_{mi}}{\eta}\right)^{\beta-1}\frac{\beta}{\eta}\exp\left\{-\left(\frac{t_{mi}}{\eta}\right)^{\beta}\right\}\right] \prod_{i=n_f+1}^{n}\left[\exp\left\{-\left(\frac{t_{mi}}{\eta}\right)^{\beta}\right\}\right]d\beta d\eta ~ d\mu < \infty. \nonumber
\end{eqnarray}


%
	
\clearpage

\newpage

\section*{References}
\bibliographystyle{natbib}

\end{document}